\documentstyle[12pt]{elsart}
  
\def\be{\begin{equation}}
\def\ee{\end{equation}}
\def\ben{\begin{displaymath}}
\def\een{\end{displaymath}}
\def\ba{\begin{array}{c}}
\def\ba2{\begin{array}{cc}}
\def\ea{\end{array}}
\def\bea{\begin{eqnarray}}
\def\eea{\end{eqnarray}}
\begin{document}


\begin{center}{\Large \bf
Construction of ${\mathcal PT}-$asymmetric non-Hermitian Hamiltonians
with ${\mathcal CPT}$ symmetry }\end{center}

\vspace{10mm}

\begin{center}
{\bf
 Emanuela Caliceti}


Dipartimento di Matematica dell'Universit\`{a} and
Istituto Nazionale di Fisica Nucleare,
I-40127 Bologna, Italy

e-mail: caliceti@dm.unibo.it

 \vspace{3mm}
{\bf Francesco Cannata}


Dipartimento di Fisica dell'Universit\`a and
         Istituto Nazionale di Fisica Nucleare,
         I-40126 Bologna, Italy

e-mail: cannata@bo.infn.it

 \vspace{3mm}
{\bf Miloslav Znojil}


\'{U}stav jadern\'e fyziky AV \v{C}R, 250 68 \v{R}e\v{z}, Czech
Republic

e-mail: znojil@ujf.cas.cz

 \vspace{3mm}
and
\vspace{3mm}

{\bf Alberto Ventura}


 Ente Nuove Tecnologie, Energia e Ambiente, and
         Istituto Nazionale di Fisica Nucleare,
         I-40126 Bologna, Italy

e-mail: ventura@bologna.enea.it

\end{center}

\vspace{5mm}

\newpage

\section*{Abstract}

Within ${\mathcal CPT}-$symmetric quantum mechanics the most
elementary differential form of the ``charge operator'' ${\mathcal C}$
is assumed. A closed-form integrability of the related coupled
differential self-consistency conditions and a
natural embedding of the Hamiltonians in a supersymmetric scheme
is achieved.  For a particular choice of the interactions the
rigorous mathematical consistency of the construction is
scrutinized suggesting that quantum systems with non-self-adjoint
Hamiltonians may admit probabilistic interpretation even in presence of a
manifest breakdown of both ${\mathcal T}$ symmetry (i.e.,
Hermiticity) and ${\mathcal PT}$ symmetry.

\vspace{5mm}

PACS 02.30.Tb  03.65Ca  03.65.Db  03.65.Ge

\newpage

\section{Introduction}

The popularity of anharmonic-oscillator models,
such as
 \ben
 H^{(AHO)}(f,g) = -{d^2 \over {dx^2}}+x^2+f\,x^3+g\,x^4
 \een
seems to reflect a fortunate combination of physical appeal
(the potential is safely confining at $g>0$) and
computational tractability. In this letter, we intend to join the
 effort of studying these models in a non-self-adjoint
regime \cite{AHO}.
While the construction of the solutions becomes
fairly easy in perturbative framework \cite{Flugge,K},
 a certain paradox  arises because
 the perturbative power series (near $f=0$)
 \ben
 E^{(AHO)}(f,g) = \sum_{k=0}^\infty f^k E^{(AHO)}_k(0,g)
 \een
 should represent the energies for  all the complex
couplings which lie in a sufficiently small circle
of convergence. A deeper analysis
 \cite{BW,RS} revealed that  the
energies $E_n^{(AHO)}(f,g)$ should  be considered as the
 infinite sets of the values of  a single analytic
function of the couplings on  different Riemann sheets.

The latter idea has steadily stimulated interest in the
manifestly non-Hermitian anharmonic oscillators~ \cite{Calic,Alv,BG}.  Finally, a real boom
of interest in similar models arose after the seminal
letter \cite{BB} by Bender and Boettcher, who argued that the
reality of the spectra should be related to the symmetries of the
Hamiltonians. Indeed, once we re-write
 Hermiticity, $H = H^\dagger$, in the form of an
involutive time-reversal symmetry, ${\mathcal T}$~\cite{erratum},
 \be
 {\mathcal T}\,H = H\,{\mathcal T} \,,
 \label{tsy}
 \ee
it is quite natural to
replace eq. (\ref{tsy}) with the constraint
 \be
 {\mathcal PT}\,H = H\,{\mathcal PT} \,
 \label{ptsy}
 \ee
where ${\mathcal P}$ denotes parity. Eq. (\ref{ptsy}) is valid for Hamiltonians
that are invariant under $\mathcal{PT}$, but not necessarily under $\mathcal{P}$
and $\mathcal{T}$ separately.

The expected reality of the energies  $E_n^{(AHO)}(i\lambda,g)$,
with real $\lambda$ and $g$, has been supported, in some cases,
by rigorous proofs \cite{DDT,Sh}.  A further
weakening of the standard Hermiticity is possible once we replace
${\cal P}$ in Eq. (\ref{ptsy}) by any other Hermitian
 operator ${\mathcal F}={\mathcal F}^\dagger$ \cite{Dirac}.
 One
thus has the new condition~\cite{AM1}

\be
{\mathcal FT}H=H{\mathcal FT} \iff {\mathcal F}H^{\dagger}=H{\mathcal F},
\label{obpsy}
\ee
which, for ${\mathcal F}= 1$ , becomes Hermiticity and for ${\mathcal F}= {\mathcal P}$
becomes ${\mathcal PT}$ symmetry.
Equation (\ref{obpsy}) implies that if $H$ has an eigenvalue $E$, then $E^*$,
apart from normalization problems, is also an eigenvalue
unless ${\mathcal F} \psi^* = 0$, so eigenvalues are either real, or enter in complex conjugate pairs~\cite{Ali}. 

One could, generically, construct many operators ${\mathcal F}$ which
would be, via eq. (\ref{obpsy}), compatible with a given
Hamiltonian $H$. Among all the possible choices of these (metric)
operators, a privileged position is occupied by the positive bounded
Hermitian ones, because the corresponding Hamiltonians admit
probabilistic interpretation~\cite{Geyer}.

\section{ ${\mathcal CPT}-$symmetric models}

\subsection{Factorized ${\mathcal F}$}

One may demand a factorization of the operator ${\mathcal
F}={\mathcal F}^\dagger$ into a product, say,  ${\mathcal F}
\equiv {\mathcal CP}$ where, conventionally, ${\mathcal C}$ can be
called a ``charge conjugation" operator~\cite{AHO,BBJ}. In
principle, this would constitute a ${\mathcal CPT}$-symmetric
quantum mechanics, with an obvious ambition of being a
zero-dimensional ${\mathcal CPT}$-symmetric field theory.

For our purposes, however, the involutory property ${\mathcal C}^2 =1$ of
the charge, or of the so called quasi-parity~\cite{ptho}, in some exactly
solvable one-dimensional examples
\be
 H =- {d^2 \over {dx^2}} +V(x),\qquad x \in {\mathbf R}
 \label{oneb}
 \ee
 is by far not necessary.
 Our interest will naturally be focused on the
possibility that any system in ${\mathcal CPT}$-symmetric quantum
mechanics  may violate  both the ${\mathcal PT}$ and
${\mathcal T}$ symmetries.

 We proceed constructively and, for the sake of
definiteness, we select the class of operators
 \ben
 {\mathcal C} = \frac{d}{dx} + w(x)
 \een
in conjunction with the above Hamiltonians (\ref{oneb}).

In our specific Ansatz for ${\mathcal C}$, in order to enforce Hermiticity
of  ${\mathcal F}$, keeping in mind that ${\mathcal P} =
{\mathcal P}^\dagger$ and ${\mathcal P}^2=1$,  we have to postulate
 definite spatial symmetry properties of the complex function
$w(x)= \sigma(x)+i\, \alpha(x)$ with
 \be
 {\mathcal R}e \ w(x) = \sigma(x) = \sigma(-x), \ \ \ \ \
 {\mathcal I}m \ w(x) = \alpha(x) = -\alpha(-x), \ \ \ \ \ \
  \ x \in {\mathbf R}\,
  \label{jed}
 \ee

\subsection{Compatibility conditions and their integrability}

Our Hamiltonian, $H$, is assumed compatible with the ${\mathcal
CPT}$ symmetry condition, i.e. Eq. (\ref{obpsy}) with ${\mathcal F}=
{\mathcal CP}$.
A verification of this condition will be the core of our present
construction. It necessitates  a decomposition of our Hamiltonian in the sum
 \be
 H = -\frac{d^2}{dx^2} + \Sigma(x) + K(x) + i\,S(x) + i\,D(x),
\label{CPTH}
 \ee
where the separate components of the complex potential $V(x)$ may
be chosen to exhibit definite parities,
 \ben
 \Sigma(x)= \Sigma(-x), \ \ \ \ \  K(x) = -K(-x), \ \ \ \ \
 S(x)= S(-x), \ \ \ \ \  D(x) = -D(-x)\,.
 \een
This simplifies the form of $H^\dagger$ and our main
compatibility constraint (\ref{obpsy}). In
principle, it should be a linear differential operator of the
third order but once we re-write it in the form of a product,
$[H\,{\mathcal C} -{\mathcal C}\,( {\mathcal P} H^\dagger\,
{\mathcal P\,)}]\,{\mathcal
P}$, we immediately see that the coefficients of the third and of
the second derivative are identically zero. A condition of the
vanishing of the coefficient of the first derivative remains
nontrivial and relates the potentials $K$ and $S$ to the choice of
${\mathcal C}$,
  \bea
  K(x)  = \frac{d}{dx}\,\sigma(x), \qquad
  S(x) & = & \frac{d}{dx}\,\alpha(x).
  \eea
In this way we are left with the condition which connects the two
complex functions $V(x)$ and $w(x)$. Its separation into  real
and imaginary part proves encouragingly simple 
 \bea
 \frac{d}{dx}\,\Sigma(x)&=& 2\,\sigma(x)\,\frac{d}{dx}\,\sigma(x) -
 2\,\alpha(x)\,\frac{d}{dx}\,\alpha(x)\,,\\
 \frac{d}{dx}\,D(x)&=& 2\,\sigma(x)\,\frac{d}{dx}\,\alpha(x) +
 2\,\alpha(x)\,\frac{d}{dx}\,\sigma(x)\,,\nonumber
 \eea
luckily admitting an entirely elementary integration, with
just a single real integration constant, $\omega$
 \bea
 \Sigma(x)= \sigma^2(x) - \alpha^2(x) + \omega\, ,\qquad
  D(x) & = & 2\,\sigma(x)\,\alpha(x)\,,\label{previ}
 \eea
 This
means that we may contemplate a family of anharmonic-oscillator
examples with $\sigma(x) = \sigma_n(x)= \mu_nx^{2n}$ and
$\alpha(x) = \alpha_n(x)= \nu_nx^{2n-1}$ for illustration, with
real $\mu_n,\ \nu_n$ and with any choice of the index $n =1,2
\ldots$.

\section{Discussion}

\subsection{Supersymmetric picture}

One can rewrite equations (\ref{previ})
supersymmetrically~\cite{ACDI}. It is not difficult to check that
$H$ of eq. (\ref{CPTH}) with $\omega =0$ satisfies
\be
H = {\mathcal F F^*}, \qquad H^\dagger = {\mathcal F^*F}, \ee
where ${\mathcal F}^*$ is the complex conjugate of ${\mathcal F}$
. We introduce the super-charges
 \bea
 Q  \equiv  \left( \ba2 0 &
{\mathcal F} \\ 0 &  0 \ea \right), \qquad
 \tilde Q & \equiv &
\left( \ba2 0 & 0 \\ {\mathcal F}^* & 0 \ea \right).
 \eea
 It is
easy to check that $Q \not= \tilde Q^\dagger$, insofar as
${\mathcal F}^\dagger \not= {\mathcal F}^*$, $Q^2 =\tilde Q^2=0$,
and $\left[ Q,H \right]=\left[ \tilde Q, H \right]=0$. The
super-Hamiltonian reads
\be
{\mathcal H} \equiv \left\{ Q,\tilde Q \right\} = \left( \ba2
{\mathcal F}{\mathcal F}^* & 0 \\ 0 & {\mathcal F}^*{\mathcal F}
\ea \right)\,.
 \ee
It is worth noting that in a way characteristic for non-Hermitian
supersymmetric examples \cite{susyho} the operator ${\mathcal H}$
is not necessarily positive.

\subsection{The problem of invertibility}

Whenever our operator ${\mathcal F}$ is unbounded~\cite{Geyer}, it may still
 be  invertible and the inverse operator may be bounded. If ${\mathcal F}^{-1}$
exists, one can derive algebraically the following relation
\be
H^{\dagger}{\mathcal F}^{-1}={\mathcal F}^{-1}H.
\label{inverse}
\ee
Supersymmetrically, one can define new conserved charges
\bea
Q^{-1}  \equiv  \left(
\ba2
0 & 0 \\
{\mathcal F}^{-1} &  0
\ea \right), \qquad
\tilde Q^{-1} & \equiv & \left(
\ba2
0 &{{\mathcal F}^*}^{-1} \\
0  & 0
\ea \right).
\eea
such that
\be
\left[ Q^{-1},H \right]=\left[\tilde Q^{-1}, H \right]=0,
\ee
with $Q^{-1}$ being not the standard inverse operator, but satisfying
\be
\left\{ Q, Q^{-1}\right\}=I\!\!I, \qquad \left\{
 \tilde Q,\tilde Q^{-1}\right\} =I\!\!I,
\ee
with $I\!\!I$ the identity operator, in analogy with the
anticommutation relations of fermion creators and annihilators.

Let us now examine in more detail the case $n=1$, with
$ \sigma (x) = {\mu}_1  x^2$, where $\mu_1 \ne 0$, and 
$\alpha (x) ={\nu}_1 x$.
${\mathcal F}$ is not bounded and not positive; however, it is invertible in $L^2({\mathbf R})$ and
Eq. (\ref{inverse}) holds. In fact, as an operator in $L^2({\mathbf R})$, ${\mathcal C}=
d/dx +\mu_1 x^2 +i \nu_1 x$ is unitarily equivalent to ${\mathcal C}_1 = d/dx + \mu_1 x^2 + \nu_1^2/(4\mu_1)$,
via the translation $x \to x - i\nu_1/(2\mu_1)$. In turn, ${\mathcal C}_1$ is unitarily
equivalent to ${\mathcal C}_2=-\mu_1 d^2/dx^2 +ix + \nu_1^2/(4\mu_1)$, via the Fourier
transformation. As is well known~\cite{He79}, ${\mathcal C}_2$ has empty spectrum and
is thus invertible; as a consequence, ${\mathcal C}$ is invertible, too; the same holds for ${\mathcal F}=
{\mathcal CP}$ (see also Ref.~\cite{Cal04}), which is therefore invertible in  $L^2({\mathbf R})$
with bounded inverse, ${\mathcal F}^{-1}$, defined on the whole $L^2({\mathbf R})$.

\subsection{The problem of positivity}

While ${\mathcal F}^{-1}$ for $n=1$ is a bounded Hermitian operator
acting on $L^2({\mathbf R})$ ~\cite{Cal04}, the positivity
requirement presents problems, in general; however,
 evaluation of matrix elements of Eq. (\ref{inverse}) between eigenstates
of the Hermitian operator ${\mathcal F}^{-1}$ yields
\bea
\left \langle j \left\bracevert H^\dagger {\mathcal F}^{-1}
 \right\bracevert k \right \rangle & = &
\left \langle j \left\bracevert {\mathcal F}^{-1}  H
 \right\bracevert k \right \rangle, \\
\lambda_k \left \langle j \left\bracevert H_1^\dagger
 \right\bracevert k \right \rangle & = &
\lambda_j \left \langle j \left\bracevert H_1
 \right\bracevert k \right \rangle \,,\label{lambda_j_k}
\eea
where $H_1$ reads
\bea
H_1 & = & - {d^2 \over {dx^2}}+ \mu_1^2 x^4 -\nu_1^2 x^2 +2
 \mu_1 x + i\nu_1 +2i\mu_1 \nu_1 x^3 \label{H_1} \\
    & \equiv & H_R +i\nu_1+2i\mu_1 \nu_1 x^3  \nonumber \;
      = \; H_R +i\nu_1V_I\,.\nonumber
\eea
We can rewrite Eq. (\ref{lambda_j_k}) as
\be
\lambda_k \left( H_R^{jk}-i\nu_1V_I^{jk}\right)=\lambda_j \left(
H_R^{jk}+i\nu_1 V_I^{jk} \right)\,, \label{lambda2} \ee where
$H_R^{jk}=A+iB$ and $V_I^{jk}=C+iD$ are complex numbers. Thus, by
equating real and imaginary parts of both sides of Eq.
(\ref{lambda2}), one gets
\be
{\lambda_k \over \lambda_j } = { {A^2 -\nu_1^2 D^2} \over
{(A+\nu_1D)^2 }} = { {B^2-\nu_1C^2}\over {(B-\nu_1C)^2}}\,,
 \ee
and, if $ \lambda_k / \lambda_j < 0$, one can argue that the
$H_R^{jk}$'s are strongly suppressed for small values of $\nu_1$.
This may lead to a practical decoupling of the two sectors
of positive and negative eigenvalues, thus supporting $ {\mathcal F}^{-1}$
as a metric operator candidate, since, physically, it is not so
important that ${\mathcal{F}}^{-1}$ is positive, but it is crucial
that the Hamiltonian connects only weakly the sectors of positive
and negative eigenvalues.

Coming now to the properties of Hamiltonian (\ref{H_1}), one can
separate the ${ \mathcal PT}$-symmetric and antisymmetric parts as
\be
H_1 = H_1^{{\mathcal PT}} + 2 \mu_1 x +i\nu_1\,,
\label{h_1}
\ee
where
\be
H_1^{{\mathcal PT}} = - { d^2 \over {dx^2}} +\mu_1^2 x^4 -\nu_1^2 x^2
 + 2i\mu_1\nu_1 x ^3\,.
 \ee
$H_1^{\mathcal PT}$ is well controlled from a mathematical point
of view, so that our proposal opens a way to study some additional
Hamiltonians enriching the class of the recent popular
non-Hermitian versions by addition of the non-${\mathcal
PT}$-symmetric Stark-like term. It is worthwhile to point out that,
performing a shift $x\rightarrow x-i\nu_1/(2\mu_1)$, one can show
that Hamiltonian (\ref{h_1}) has real spectrum~\cite{Ba04}.

In general, for all $\mu,\nu \in \mathbf R$, let $H(\mu, \nu)$
denote the Schr\"odinger operator in $L^2 (\mathbf R)$ defined by
\begin{equation}
H(\mu,\nu)
= - { d^2 \over {dx^2}} +\mu ^2 x^4 +2i\mu \nu x^3 -\nu ^2 x^2 +2\mu x
 \end{equation}
on the domain $D = H^2 ({\mathbf R})\cap L^2_4({\mathbf R})$. Then
$H(\mu,\nu)$ is a closed operator with compact resolvents and,
therefore, discrete spectrum. In fact, the operator
$-\frac{d^2}{dx^2}+\mu ^2 x^4$ enjoys such properties (see
\cite{S}), which extend to the analytic family 
\begin{equation}
H_g (\mu,\nu)=-\frac{d^2}{dx^2}+\mu ^2 x^4 +g(2i\mu \nu x^3 -\nu ^2
x^2 +2\mu x)
\end{equation}
for $g\in \mathbf C$ and in particular to the original operator
$H(\mu,\nu)$ for $g=1$ (for more details on the theory of
analytic families of operators see \cite{K} or \cite{RS}).

If we now introduce a further perturbation parameter $\gamma \in
\mathbf C$ only in the linear term:
\begin{equation}
H_{\gamma}(\mu,\nu)=-\frac{d^2}{dx^2}+\mu ^2 x^4 +2i\mu \nu x^3 -\nu
^2 x^2 +2\mu \gamma x,
\end{equation}
then $H_{\gamma=0} (\mu,\nu)$ is $\cal{PT}$-symmetric with real spectrum
\cite{Sh}, while for finite non-zero values of $\gamma$ the spectrum is
complex~\cite{Ba04}.

The spectral analysis for the complete operator $H_{\gamma}(\mu,\nu)$
for $\gamma \in \mathbf R$ can be performed in the framework of
perturbation theory around $\gamma =0$. More precisely, referring to
 results in Ref.~\cite{CGS}, it is possible to
prove that for fixed $\mu$ and $\nu$ there exists $\delta >0$ such that
the eigenvalues of
$H_{\gamma}(\mu,\nu)$ are real and represent a sequence of analytic
functions $E_n (\gamma)$ for $\gamma \in ]-\delta,\delta[$. For such values
of
$\gamma$ each eigenvalue $E_n (\gamma)$ is the sum of the
corresponding Rayleigh-Schr\"odinger perturbation expansion around
the eigenvalue $E_n (0)$ of $H_0 (\mu,\nu)$.

Yet to be explored is the usefulness of  second and higher order
 derivatives in the ansatz for the ${\mathcal C}$ operator, with the
possibility of non-linear algebraic structures~\cite{Ba04,AAA}.

\section*{Acknowledgements}

The authors wish to thank Prof. S. Graffi and Prof. B. Bagchi for
useful discussions and for reading the manuscript. M. Z.
appreciates the hospitality of Dipartimento di Matematica,
Universit\`{a} di Bologna where this work has been initiated.
Partially, his participation was also supported by  GA AS {C}R,
grant Nr. 104 8302 and by AS CR project AV0Z1048901.

\end{document}